\documentclass[twocolumn]{aastex701}

\usepackage[colorinlistoftodos]{todonotes}
\usepackage{amssymb}
\usepackage{comment}
\usepackage{multirow,acronym}
\usepackage{natbib}
\usepackage{makecell} 
\usepackage{booktabs}
\usepackage{amsmath}
\usepackage{subfigure}
\usepackage{color}
\usepackage{xcolor}
\usepackage{hyperref}
\usepackage[T1]{fontenc}
\usepackage{graphicx}
\usepackage{bm}
\newcommand{\figdir}{.}

\begin{document}

\title{From Inspiral to Expansion: The Wake-Driven Torque on Binary Black Holes in Gaseous Medium}

\author[0009-0006-7551-6433]{Jixuan Yang}
\affiliation{Department of Astronomy, Tsinghua University,
Beijing 100084, China}
\email{yang-jx24@mails.tsinghua.edu.cn}

\author[0000-0002-6540-7042]{Lile Wang}
\affiliation{The Kavli Institute for Astronomy and Astrophysics, Peking
University, Beijing 100871, China} \affiliation{Department
of Astronomy, School of Physics, Peking University, Beijing
100871, China} 
\email{lilew@pku.edu.cn}

\author[0000-0003-0750-3543]{Xinyu Li}
\affiliation{Department of Astronomy, Tsinghua University,
Beijing 100084, China} 
\email{xinyuli@tsinghua.edu.cn}

\author[0000-0001-9222-4367]{Rixin Li}
\affiliation{Department of Astronomy, University of California Berkeley,
Berkeley, CA 94720-3411, USA} 
\email{rixin@berkeley.edu}

\correspondingauthor{Xinyu Li} \email{xinyuli@tsinghua.edu.cn}

\correspondingauthor{Lile Wang} \email{lilew@pku.edu.cn}
\begin{abstract} 

Binary black holes (BBHs) in gaseous medium, such as active galactic nucleus (AGN) disks, are important gravitational-wave sources, yet the gas-driven torque that governs their orbital evolution remains to be fully understood. 
Most existing studies of BBHs in gas often approximate the net torque as the sum of independent dynamical friction (DF) forces exerted on each black hole by its own wake, neglecting the mutual gravitational coupling between the two wakes.
We perform three-dimensional hydrodynamic simulations of circular BBHs in a uniform flow and find the torque is determined by the wake-wake interactions.
The net torque is controlled by a single parameter $\eta \equiv v_g/v_o$, the ratio of the gas flow velocity to the binary orbital velocity. 
At small $\eta$, the wakes merge into a single overdense envelope and the time-averaged torque is negative; as $\eta$ increases, the wakes separate and the torque becomes positive.
In AGN disks, capture-channel binaries in a disk model naturally produce $\eta$ in the positive-torque regime; the expansion timescale is comparable to or shorter than the disk lifetime, suggesting that gas-driven expansion can compete with gravitational-wave inspiral and suppress the capture-channel merger rate.

%We perform three-dimensional hydrodynamic simulations of binary black holes (BBHs) moving through a uniform gaseous medium. The gas morphology and net torque depend primarily on a single dimensionless parameter $\eta \equiv v_g/v_o$, the ratio of the gas flow velocity to the binary orbital velocity. At small $\eta$ the wakes merge into a common overdense envelope and the time-averaged torque is negative; as $\eta$ increases, the wakes separate and the torque becomes positive. The torque decays as a power law in $\eta$ at large values, scales linearly with mass ratio, and reverses sign near an intermediate orbital inclination, with approximately half of isotropic inner orbits yielding positive torque. A single orbiting particle reproduces the Ostriker dynamical friction prediction, whereas the binary torque deviates substantially, demonstrating that wake--wake and wake--BH interactions , rather than independent DF, produce the torque. Applied to AGN disks, capture-channel binaries in a disk model naturally produce $\eta$ in the positive-torque regime; the expansion timescale is comparable to or shorter than AGN disk lifetimes, suggesting that gas-driven expansion can compete with gravitational-wave inspiral and suppress the capture-channel merger rate.

\end{abstract}


\section{Introduction}

Binary black holes (BBHs) constitute the primary sources of gravitational waves (GWs) detected by the LIGO--Virgo--KAGRA collaboration, and understanding their orbital evolution is essential for predicting merger rates and interpreting GW observations. In vacuum, orbital decay can be driven by gravitational-wave emission \citep{1964PhRv..136.1224P}, dynamical encounters in dense star clusters, or Kozai--Lidov oscillations \citep{1962AJ.....67..591K,1962P&SS....9..719L} in triple systems. In many astrophysical environments, however, BBHs are embedded in gaseous media, including active galactic nuclei (AGN) disks, common-envelope systems \citep{1976IAUS...73...75P}, and gas-rich young star clusters \citep{2016MNRAS.459.3432M}, and the interaction with ambient gas can exchange energy and angular momentum, significantly altering the orbital evolution. \citet{2020ApJ...898...25T} predicted that the AGN channel alone may contribute a merger rate of $\sim 0.02$--$60~{\rm Gpc^{-3}~yr^{-1}}$, a substantial fraction of the total rate inferred from LIGO observations ($\sim 13$--$26~{\rm Gpc^{-3}~yr^{-1}}$; \citealt{2026ApJ..1005L..51A}). Determining how gas affects BBH dynamics is therefore crucial for distinguishing between formation channels and interpreting the observed GW event rate.

The most widely studied gas-related mechanism is gaseous dynamical friction (DF): a gravitating object moving through gas excites a density wake behind, and the gravitational attraction from this overdense wake produces a drag force opposite to the direction of motion. Building on the classical dynamical friction theory of \citet{1943ApJ....97..255C}, \citet{1999ApJ...513..252O} derived the corresponding expression for a gaseous medium using linear perturbation theory, distinguishing between the supersonic regime (wake confined within a Mach cone) and the subsonic regime (nearly spherical perturbation). Subsequent work extended gaseous DF to orbital motion and nonlinear gas response \citep{2007ApJ...665..432K,2009ApJ...703.1278K}, and \citet{2008ApJ...679L..33K} investigated DF on double perturbers in a gaseous medium. \citet{2019ApJ...884...22A} applied this framework to binary Bondi--Hoyle--Lyttleton (BHL) accretion. These treatments, however, model each perturber independently and do not account for gravitational coupling between the wakes generated by different objects.

\begin{figure}
    \centering
    \includegraphics[width=1\linewidth]{\figdir/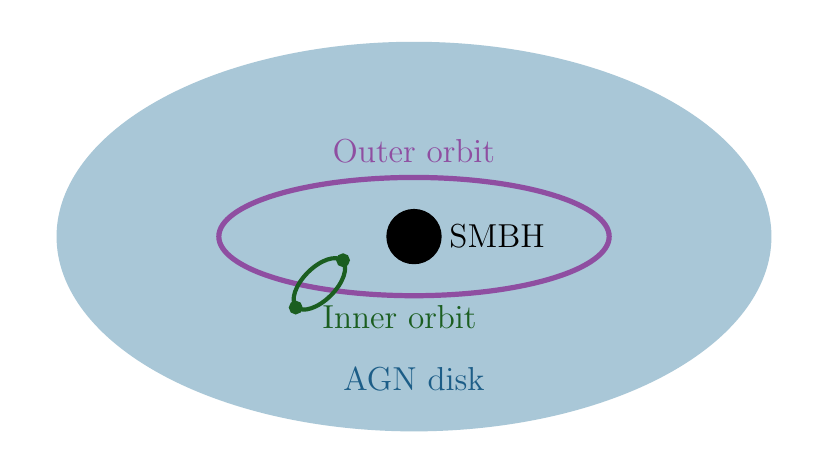}
    \caption{Schematic diagram of the inner binary orbit and the outer orbit around the SMBH.}
    \label{fig:inner_outer_orbit}
\end{figure}

For BBHs embedded in AGN disks, the binary's inner orbit (the two components about their center of mass) is nested within a much larger outer orbit (the center of mass about the central SMBH). This configuration is sketched in Figure~\ref{fig:inner_outer_orbit}.
Numerous numerical studies have employed circum-single disk (CSD) simulations: \citet{2011ApJ...726...28B} found that gas torques can harden gap-opening, intermediate-mass BBHs, \citet{2021ApJ...911..124L} showed better resolved CSDs reverse the torque sign; \citet{2022MNRAS.517.1602L,2023MNRAS.522.1881L,2024MNRAS.529..348L} further examined gas-driven hardening of stellar-mass BBHs, demonstrating its dependence on disk and BBH parameters, \citet{2022ApJ...940..155D}, using 3D shearing-box simulations, showed that the torque can change sign as the binary separation varies.
%For BBHs embedded in AGN disks, numerous numerical studies have employed circum-single disk (CSD) simulations: \citet{2021ApJ...911..124L} found that gas torques can drive binary hardening, \citet{2022MNRAS.517.1602L,2023MNRAS.522.1881L,2024MNRAS.529..348L} examined the torque dependence on disk parameters, \citet{2022ApJ...940..155D} showed using 3D shearing-box simulations that the torque sign can reverse with binary separation, and \citet{2024ApJ...970..107C} demonstrated that retrograde binaries can experience runaway eccentricity growth.
% outflow--ambient interaction studies \citep{2020MNRAS.492.2755G,2020MNRAS.494.2327L,2025ApJ...992....7L} have shown that the sign and magnitude of gas-induced forces also depend on the surrounding gas structure. % TODO
These CSD-based simulations share a common assumption: the ratio of the velocity of the binary center of mass to the orbital speed of each black hole is low, so that individual CSDs can form around each black hole in most cases.\footnote{CSD formation can still be suppressed when the relative velocity between each black hole and the gas is high, as in retrograde binaries.}
%so that individual CSDs can form around each black hole.
In many astrophysical scenarios, however, the relative velocity can be substantial; for instance, when the binary follows an eccentric or inclined orbit around the central supermassive black hole (SMBH), or when the disk gas deviates from Keplerian motion due to accretion-driven radial inflow, and in this regime CSDs cannot form; instead, each black hole (BH) generates a Bondi--Hoyle--Lyttleton (BHL) wake. 
%The degree of wake overlap is governed by the ratio of the orbital separation to the BHL radius. When the ratio is large, the BHL spheres are well separated and each BH generates its own distinct wake. 
This regime has not been systematically investigated. 

To study how the gravitational torque from the wake--wake and wake--BH interactions may deviate from the independent-DF picture, we perform three-dimensional hydrodynamic simulations of a passive, circular black hole binary embedded in a uniform gaseous flow, as the binary separation is much smaller than the characteristic length scales of the accretion disk.
We neglect the gravitational influence of the central SMBH on the BBH orbit and assume no winds or outflows are launched from the black holes. 

This paper is structured as follows. In Section~\ref{sec:methods}, we introduce the numerical methods, simulation setup, and model parameters. In Section~\ref{sec:fiducial}, we present the fiducial model, including gas density distribution, drag force, and torque. In Section~\ref{sec:parameter}, we investigate the dependence of the results on the binary and gas parameters. In Section~\ref{sec:discussion}, we 
%interpret the physical mechanism of the wake-driven torque (\S\ref{sec:physical_mechanism}), 
compare our results to prior numerical studies, discuss the implications for binary black hole mergers in AGN disks (\S\ref{sec:disc_agn}), and address the principal caveats and promising directions for future work (\S\ref{sec:caveats}). Appendices~\ref{apx:analyse} and~\ref{apx:convergence} provide the analytic dynamical-friction comparison and resolution convergence tests, respectively. We summarize our principal conclusions in Section \ref{sec:conclusion}.

\section{Methods}\label{sec:methods}
\begin{figure}
    \centering
    \includegraphics[width=.8\linewidth]{\figdir/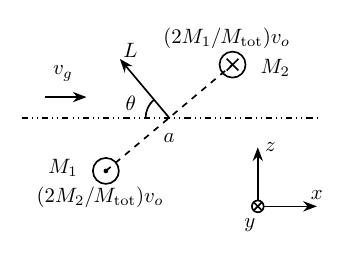}
    \caption{Schematic illustration of the simulation setup in the $x$--$z$ plane. The binary components with masses $M_1$ and $M_2$ are separated by $a$ and orbit about the center of mass with velocities $(2M_2/M_{\rm tot})v_o$ and $(2M_1/M_{\rm tot})v_o$, respectively, where $v_o$ is the average orbital velocity. The ambient gas flows uniformly at velocity $v_g$. The inclination angle $\theta$ between the binary's orbital angular momentum $\mathbf{L}_{\rm orb}$ and the $-x$ direction is indicated. The dashed line marks the direction of $v_g$ (the $x$ axis).}
    \label{fig:setup_schematic}
\end{figure}

We use the GPU-dedicated code \texttt{Kratos} \citep{2025ApJS..277...63W} to perform numerical simulations. 
The simulation setup is illustrated schematically in Figure~\ref{fig:setup_schematic}.
In the center-of-mass frame of the binary, two BHs with mass $M_1$ and $M_2$ are moving on the circular orbits of radius $a$.
The binary center of mass is fixed at the center of the computational box. 
%Unless otherwise specified, the binary orbit lies in the $x$--$y$ plane, corresponding to the inner orbit illustrated in Figure~\ref{fig:inner_outer_orbit}. 
At large distances from the binary, the gas has uniform density and flows along the $+x$ direction, with inclination angle $\theta$ relative to the orbital angular momentum of the BBH.

Throughout Section \ref{sec:methods}--Section \ref{sec:parameter} we work in code units. The total mass of the binary is set to $GM_{\rm tot}=1$ where $M_{\rm tot}=M_1+M_2$, the orbital separation is $a=1$. The gas is modeled as an ideal gas with adiabatic index $\gamma=5/3$. We adopt a uniform initial pressure of $p_0=1.5$, corresponding to a sound speed of $c_s=0.5$. All models considered in this work are supersonic: the gas flow velocity satisfies $v_g> c_s$.

\subsection{Characteristic Scales}
\label{sec:scales}

The characteristic scale governing gas accretion onto each black hole is the Bondi radius
\begin{equation}
    r_B = \frac{2GM}{c_s^2+v_*^2},
    \label{eq:rB}
\end{equation}
where $c_s$ is the sound speed of the ambient gas and $v_*$ is the relative velocity between the black hole and the surrounding gas. Considering an equal-mass binary, $M$ denotes the mass of each BH. Physically, $r_B$ is the distance at which the gravitational potential energy of the black hole equals the sum of the thermal and kinetic energy of the gas; within this radius the gas is gravitationally bound and can accrete. The Bondi radius therefore sets the minimum spatial scale that simulations must resolve, and the domain must extend well beyond $r_B$ to capture the full accretion flow. The characteristic timescale for this accretion process is
\begin{equation}
    t_B = \frac{2GM}{(c_s^2+v_*^2)^{3/2}},
    \label{eq:tB}
\end{equation}
which represents the free-fall timescale from the Bondi radius; the total simulation time should satisfy $t_{\rm tot}\gg t_B$ to ensure the flow reaches a steady state. In the high-Mach-number limit ($v_*\gg c_s$), the thermal contribution is negligible and the Bondi radius simplifies to $r_B\sim 2GM/v_*^2$. For the binary, $v_*$ in the center-of-mass frame is dominated by the ambient gas velocity $v_g$, so $v_*\sim v_g$. This approximation is justified for all models considered in this work, for which the gas Mach number $\mathcal{M}\equiv v_g/c_s\ge2$.

The binary itself introduces a second characteristic scale: the orbital separation. For a circular Keplerian orbit,
\begin{equation}
    a = \frac{GM}{2v_o^2},
    \label{eq:d}
\end{equation}
where $v_o$ is the orbital velocity of each component about the center of mass. Comparing the two scales,
\begin{equation}
    \frac{a}{r_B} \sim \frac{GM/2v_o^2}{2GM/v_g^2}
    = \frac{v_g^2}{4v_o^2}.
    \label{eq:d_over_rB}
\end{equation}
The ratio depends only on the relative velocity between the gas and the binary, suggesting that the combination $v_g/v_o$ is the natural control parameter. We therefore define
\begin{equation}
    \eta \equiv \frac{v_g}{v_o},
    \label{eq:eta}
\end{equation}
so that $a/r_B = \eta^2/4$. More generally, for $q\neq1$ the two components have different orbital velocities; we define $v_o$ as their average, $v_o \equiv \frac{1}{2}\sqrt{GM_{\rm tot}/a}$, which reduces to the equal-mass orbital velocity of each component when $q=1$.
The ratio $a/r_B$ directly measures the degree of wake overlap: when $a/r_B\ll1$ ($\eta\ll1$), the Bondi spheres of the two black holes overlap substantially and the binary is embedded within a single shared envelope; when $a/r_B\gg1$ ($\eta\gg1$), the two BHL spheres are well separated and each black hole generates its own distinct wake. The shared-envelope regime has been studied in the context of BHL accretion by \citet{2019ApJ...884...22A} at $\eta=2$; the present work explores the isolated-wake regime. We expect the wake morphology to be reflected in the net gas torque on the binary.

Other dimensionless parameters that characterize the system include the mass ratio $q\equiv M_2/M_1$, the orbital eccentricity $e$, and the inclination angle $\theta$ between the binary's angular momentum vector and $-x$ direction (shown in Figure \ref{fig:setup_schematic}). In this work we focus on circular, edge-on orbits ($e=0$, $\theta=90^{\circ}$) as the fiducial configuration; with $\eta$ identified as the primary physical parameter governing the gas morphology, we now describe the numerical implementation.

\subsection{Basic Configurations and Boundary Conditions}
\label{sec:config}

%We adopt a fixed-orbit approach: the binary orbit is held fixed throughout each simulation, with the binary separation fixed at $a=1$ in code units. 
We keep the orbital separation $a$ fixed during the simulation.
This approach isolates the hydrodynamic torque from orbital back-reaction, enabling a clean measurement of the instantaneous torque exerted by the gas on the binary as a function of the instantaneous orbital parameters. By decoupling the gas torque from orbital evolution, we avoid the complication of feedback between torque-driven orbital changes and the resulting gas morphology, allowing direct comparison with analytic torque estimates. The orbital separation is many orders of magnitude larger than the Schwarzschild radius for the modeled black hole masses, so general relativistic corrections are negligible.

The computational domain is a Cartesian cube of side length $L=32$ with a base resolution of $256^3$ grid cells. The domain size is chosen to be much larger than the Bondi radius. The largest Bondi radius across our parameter sweep occurs at the smallest gas velocity, $\eta=2$ ($v_g=1$ in code units). So the minimum relative velocity is $v_{*,\rm min}=v_g-v_o=0.5$. The Bondi radius therefore satisfies
\begin{equation}
    r_B \lesssim \frac{2GM}{v_{*,\rm min}^2+c_s^2}\ll L,
    \label{eq:domain_adequacy}
\end{equation}
where the bounding box is sufficiently large to be conservative: including the thermal contribution ($c_s=0.5$) yields $r_B\lesssim 2$ for $\eta=2$, and $r_B$ decreases rapidly with increasing $\eta$ (e.g., $r_B\lesssim 0.24$ at $\eta=5$ and $r_B\lesssim 0.01$ at $\eta=20$). The domain thus comfortably exceeds the Bondi radius for all models.

We employ static mesh refinement (SMR) around each black hole with three levels of refinement (factor of 2 per level), yielding 
%an effective resolution of $1024^3$ grid cells at the sinks and 
a minimum cell spacing of $\delta_{\min}=1/64$.
The hydrodynamic solver uses piecewise-linear method (PLM) reconstruction, an HLLC Riemann solver, and second-order Runge--Kutta (RK2) time integration. The GPU-dedicated architecture of \texttt{Kratos} \citep{2025ApJS..277...63W} makes this resolution achievable within practical simulation runtimes. Each black hole is represented by a sink particle with radius $r_{\rm sink}=0.05$, approximately $3\delta_{\min}$ ($\delta_{\min}=1/64\approx0.0156$); 
%convergence with respect to spatial resolution is verified in Appendix~\ref{apx:convergence}.
resolution convergence is demonstrated through comparison with a $2\times$ higher-resolution run (Appendix~\ref{apx:convergence}).
Within the sink region, we impose a density floor of $\rho_{\rm floor}=10^{-8}$ to prevent numerical divergence.

The total simulation time is $t_{\rm tot}=48$, which spans $\sim 8$ orbital periods and greatly exceeds the Bondi crossing time,
\begin{equation}
    t_B \lesssim \frac{2GM}{(v_{*,\rm min}^2+c_s^2)^{3/2}}\ll t_{\rm tot},
    \label{eq:ttot_adequacy}
\end{equation}
where the total simulation time is sufficiently large to be conservative: $t_B\lesssim 2.8$ at $\eta=2$. For larger $\eta$, $t_B$ drops steeply (e.g., $t_B\approx0.11$ at $\eta=5$), ensuring that the flow reaches a steady state well before the end of each simulation for all models in the parameter sweep. The combination of a large domain and a long integration time guarantees that both the global mass accretion rate and the hydrodynamic torque converge to their steady-state values.

For the fiducial edge-on configuration ($\theta=90^\circ$), the binary lies in the $x$--$y$ plane and the midplane is a symmetry plane; we therefore simulate only the $z>0$ half of the domain with a reflecting boundary at $z=0$, reducing the computational cost by a factor of two. For the inclined-orbit models ($\theta\neq90^\circ$), the midplane symmetry is broken and the full domain ($-L/2<z<L/2$) is used, with outflow conditions applied on both $z$ faces. Gas is introduced at the upstream boundary ($-x$ face) via a fixed-state inflow condition, where the primitive variables in the ghost zones are set to the prescribed ambient density, pressure, and velocity. The remaining faces ($+x$, $\pm y$, and $+z$ for the half-domain or $\pm z$ for the full domain) use outflow conditions, implemented via zero-gradient extrapolation, allowing gas to leave the domain without producing spurious reflections. Two ghost cells are used on each domain face to support the PLM reconstruction; the boundary conditions described above are applied by setting the primitive variables in these ghost zones.

\subsection{Setup of Fiducial and Other Models}
\label{sec:models}

Table~\ref{tab:sims_list} summarizes all simulation models considered in this work. 
Model names encode the key parameters: \texttt{a1} denotes the fixed binary separation $a=1$, \texttt{E$n$} denotes $\eta=n$, and optional suffixes specify non-default values of $q$ and $\theta$. For example, \texttt{\_q09} indicates $q=0.9$, and \texttt{\_t10} indicates $\theta=10^\circ$. The defaults are $q=1$ and $\theta=90^\circ$, for which no suffix is used.

The fiducial model is \texttt{a1\_E5}: a circular, equal-mass binary ($q=1$) with $\eta=5$ and $\theta=90^\circ$, corresponding to a gas flow at $v_g=5v_o$ in the $+x$ direction. The choice of equal masses and edge-on orientation eliminates geometric asymmetry and maximizes the projected cross-section for wake interactions. %At $\eta=5$, the Bondi spheres are well separated relative to the orbital scale ($a/r_B\simeq 6.25$ by Equation \ref{eq:d_over_rB}).%, placing the system in the isolated-wake regime where each black hole generates its own distinct wake.

\begin{deluxetable*}{cccccc}
    \centering
    \caption{Summary of simulation models and parameters.}
    \label{tab:sims_list}
        \tablehead{
        \colhead{Series} & \colhead{Model} & \colhead{$e$} & \colhead{$q$} & \colhead{$\eta$} & \colhead{$\theta$}}
        \startdata
        Fiducial & \texttt{a1\_E5} & 0 & 1 & 5 & $90^\circ$\\
        $\eta$ & \texttt{a1\_E2--E20} & 0 & 1 & 2--20 & $90^\circ$\\
        $q$ & \texttt{a1\_E5\_q09--q005} & 0 & 0.05--0.9 & 5 & $90^\circ$\\
        $\theta$ & \texttt{a1\_E5\_t0--t90} & 0 & 1 & 5 & $0^\circ$--$90^\circ$\\
        \enddata
        \tablecomments{The $\theta=90^{\circ}$ entry (\texttt{a1\_E5\_t90}) is the fiducial model \texttt{a1\_E5}.}
\end{deluxetable*}

To investigate the dependence on the relative velocity between the binary and the ambient gas, we perform an $\eta$-series with $\eta$ ranging from 2 to 20 while keeping $q=1$, $e=0$, and $\theta=90^\circ$ fixed. This range spans two orders of magnitude in the Bondi-radius-to-separation ratio $a/r_B\sim\eta^2/4$.%, from the common-envelope regime ($\eta=2$) to the deeply isolated-wake limit ($\eta=20$).

To probe the mass-ratio dependence, we carry out a $q$-series with $q$ ranging from 0.05 to 0.9 at fixed $\eta=5$, $e=0$, and $\theta=90^\circ$, with the total mass $M_{\rm tot}=M_1+M_2$ held fixed. To assess the effect of the inclination angle, we vary $\theta$ from $0^\circ$ to $90^\circ$ at fixed $\eta=5$, $q=1$, and $e=0$, with the $\theta=90^\circ$ case provided by the fiducial model. This series probes how the wake interaction geometry affects the torque when the binary orbit is not aligned edge-on to the flow: as $\theta$ increases, an increasing fraction of the orbital motion lies parallel to the incoming gas stream, altering the relative velocity experienced by each black hole over the course of an orbit.

All other parameters (the binary separation $a=1$, the gas thermodynamics ($\gamma=5/3$ and $p_0=1.5$), and the Cartesian grid configuration with SMR, domain size, and boundary conditions described in \S\ref{sec:config}) are held fixed across all models. The results of this simulation campaign are presented in \S\ref{sec:fiducial} and \S\ref{sec:parameter}.

\subsection{Diagnostics}
\label{sec:diagnostics}

The key quantities used to characterize the gas--binary interaction are the gravitational drag force on each black hole, the resulting torque on the binary, and time-averaged diagnostics computed over the steady-state portion of each simulation.

At each time step, the gravitational force exerted by the gas on black hole $i$ is computed by summing the gravitational acceleration over all fluid cells, excluding cells within the sink radius $r_\mathrm{sink}$ to avoid self-force divergence:
\begin{equation}
    \mathbf{F}_i = \sum_{j\in\mathrm{gas},\,|\mathbf{r}_j-\mathbf{r}_i|>r_\mathrm{sink}} \frac{G M_i \rho_j \Delta V_j}{|\mathbf{r}_j - \mathbf{r}_i|^3}\,(\mathbf{r}_j - \mathbf{r}_i),
    \label{eq:F_i}
\end{equation}
where $\rho_j$ and $\Delta V_j$ are the mass density and volume of gas cell $j$.

For edge-on orbits ($\theta=90^{\circ}$), the instantaneous torque $\mathcal{T}(t)$ exerted by the gas on the binary is the $z$-component of the total moment of these forces about the binary center of mass:
\begin{equation}
    \mathcal{T}(t) = \sum_{i=1}^{2} \bigl(\mathbf{r}_i \times \mathbf{F}_i\bigr)_z,
    \label{eq:torque_def}
\end{equation}
where $\mathbf{r}_i$ is the position vector of black hole $i$ measured from the binary center of mass. When $\theta = 90^{\circ}$, the binary orbital angular momentum is directed along $+z$, so positive $\mathcal{T}$ corresponds to an increase in the orbital angular momentum (orbital expansion) and negative $\mathcal{T}$ to orbital shrinkage.

From the torque time series, we compute the time-averaged torque over the steady-state phase:
\begin{equation}
    \overline{\mathcal{T}} = \frac{1}{t_2 - t_1} \int_{t_1}^{t_2} \mathcal{T}(t)\, dt,
    \label{eq:time_avg}
\end{equation}
where $t_1=12$ and $t_2=43.4$ in code units, spanning five orbital periods for the fiducial model after the initial transient ($t\lesssim10$) has settled. %The uncertainty is the standard deviation of the per-cycle averages evaluated over this interval.

The force and torque time series for the fiducial model, including the transient behavior and approach to steady state, are presented in Section~\ref{sec:torque}. For context, the classical gaseous dynamical friction formula of \citet{1999ApJ...513..252O} provides an analytic baseline; the derivation and application to our simulation results are given in Appendix~\ref{apx:analyse}.

\section{Fiducial Model}\label{sec:fiducial}

\subsection{Density Distribution}
\label{sec:density}

\begin{figure*}
    \centering
    \includegraphics[width=1\linewidth]{\figdir/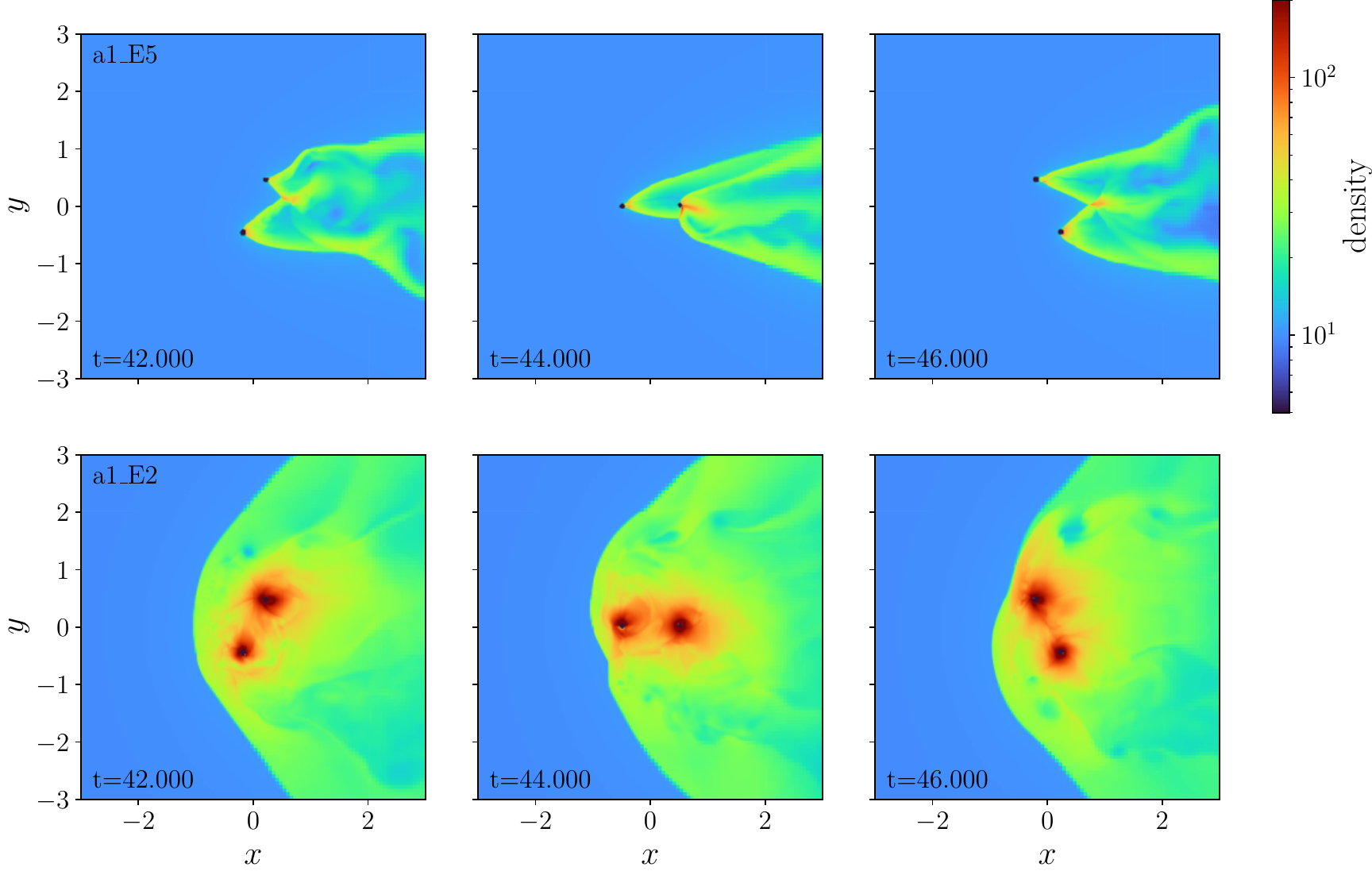}
    \caption{Gas density distributions on the $x$--$y$ plane for models \texttt{a1\_E2} and \texttt{a1\_E5} at three different orbital phases within one orbital period. The top row shows model \texttt{a1\_E5} ($\eta=5$), while the bottom row shows model \texttt{a1\_E2} ($\eta=2$).}
    \label{fig:rho_M2M5}
\end{figure*}

We examine the gas density morphology for the fiducial model \texttt{a1\_E5} ($\eta=5$) and compare it with model \texttt{a1\_E2} ($\eta=2$). As discussed in \S\ref{sec:scales}, the ratio $a/r_B=\eta^2/4$ determines the degree of wake overlap: at $\eta=2$, $d/r_B\sim 1$ and the BHL radii are comparable to the orbital separation, while at $\eta=5$, $a/r_B\sim 6.25$ and they are well separated. Figure~\ref{fig:rho_M2M5} shows the density distribution on the $x$--$y$ plane at three snapshots ($t=42,~44,~46$), with the top row showing \texttt{a1\_E5} and the bottom row \texttt{a1\_E2}. Model \texttt{a1\_E2} reproduces the results of \citet{2019ApJ...884...22A}: at $\eta=2$, the density enhancements produced by each black hole overlap substantially, embedding both black holes within a single overdense envelope that fills much of the orbital region. In this shared-envelope regime, the binary behaves approximately as a single perturber — the individual wakes cannot be distinguished as separate structures at any orbital phase, and the gravitational interaction with the gas produces a net negative torque that drives orbital contraction.

At $\eta=5$, the Bondi spheres are well separated relative to the orbital scale and each black hole generates its own narrow, high-density wake trailing in the downstream ($+x$) direction. The effective Mach number of the gas flow relative to each black hole varies over the orbit — reaching a maximum when the black hole moves upstream and a minimum when it moves downstream — but remains supersonic at all phases, so the wakes are confined to wedge-like Mach cones whose opening angle varies with the instantaneous Mach number. The density contrast between the wake interior and the ambient gas is large: the wakes appear as bright, sharply bounded structures in Figure~\ref{fig:rho_M2M5}, with peak densities exceeding the ambient value by more than an order of magnitude near each black hole. Unlike the $\eta=2$ case, the two wakes remain clearly identifiable as distinct structures for most of the orbit, with no merged envelope connecting them. At $\eta=5$, the shock front of each Mach cone attaches directly to the black hole, whereas at $\eta=2$ it is detached and stands off at a finite distance. The wakes are narrowest when a black hole moves upstream and broaden when moving downstream, producing a characteristic asymmetry over each orbital period that is not present in the merged-envelope case.

In Figure~\ref{fig:rho_M2M5}, the binary rotates counterclockwise while the wakes trail downstream at all times; at $t=44$ in the $\eta=5$ case, one black hole crosses the wake of its companion. In the snapshot of $t=46,~\eta=5$, the wake opening angle of the upper black hole ($y>0$) is smaller than that of the lower black hole ($y<0$), reflecting the phase-dependent effective Mach number. Under the sign convention of Equation \eqref{eq:torque_def} — where positive $\mathcal{T}$ corresponds to an increase in the orbital angular momentum along $+z$ — the upper wake is expected to contribute torque primarily in the $-z$ direction and the lower wake in the $+z$ direction. This geometric asymmetry % in the wake--wake interaction 
may produce a net $+z$ torque aligned with the binary angular momentum; the resulting torque asymmetry and its time evolution are quantified in \S\ref{sec:torque}. 
In summary, the $\eta=2$ case produces a merged overdense envelope yielding negative torque (inspiral), while at $\eta=5$ the wakes are narrow, isolated, and periodically cross each other's paths, producing a net positive torque (expansion) as shown in the next subsection.

\subsection{Gravitational Drag and Torque}
\label{sec:torque}

\begin{figure}
    \centering
    \includegraphics[width=1\linewidth]{\figdir/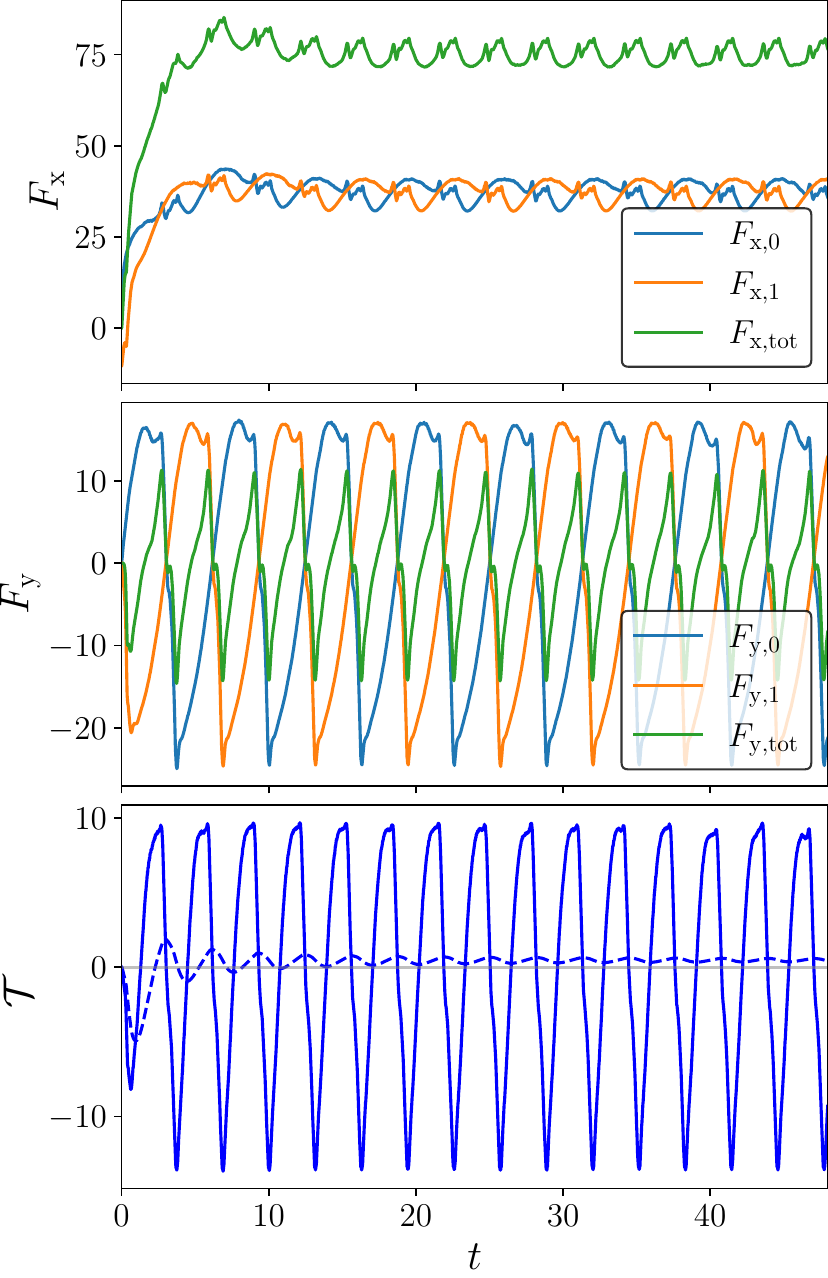}
    \caption{Time evolution of the drag force components ($F_x$, $F_y$) and torque $\mathcal{T}$ for the fiducial model \texttt{a1\_E5}. In the bottom panel, the dashed line shows the mean torque averaged from $t=0$, while the grey line marks $\mathcal{T}=0$.}
    \label{fig:curve_M5}
\end{figure}

The distinct wakes and periodic wake-crossing described above produce time-varying forces on each black hole. Figure~\ref{fig:curve_M5} shows the evolution of the drag force components $F_x$, $F_y$ (defined in Equation \eqref{eq:F_i}) and the gravitational torque $\mathcal{T}$ (Equation \eqref{eq:torque_def}) for the fiducial model \texttt{a1\_E5}. Under the sign convention of Equation \eqref{eq:torque_def}, a positive torque corresponds to an increase in the orbital angular momentum of the binary. During the early stage ($t\lesssim 10$), the drag forces build up from zero as the wakes develop from the initially uniform gas and propagate downstream. The initial transient lasts approximately two orbital periods, during which the density perturbations launched by each black hole %travel downstream and
establish the quasi-steady wake structures visible in Figure~\ref{fig:rho_M2M5}. 
%By $t\sim 10$, the transient has decayed, consistent with the steady-state criterion $t_1=12$ established in \S\ref{sec:diagnostics}. 
After $t\sim10$, the force components settle into quasi-periodic oscillations at the orbital period. The force oscillations are not purely sinusoidal, reflecting the nonlinear wake dynamics and the phase-dependent effective Mach number of each black hole relative to the flow. The total drag force $F_x$, parallel to the gas flow, oscillates about a positive mean, indicating that the gas transfers net momentum to the binary in the flow direction, as expected for gravitational drag. The perpendicular component $F_y$ arises primarily from the orbital modulation of the wake geometry and oscillates at a similar frequency, with an amplitude comparable to that of $F_x$. The individual force contributions from each black hole (shown as separate curves in Figure~\ref{fig:curve_M5}) confirm that both experience comparable forces with a phase offset of $\pi$ between them, reflecting the symmetric orbital configuration of the equal-mass binary: when one BH leads in its orbit, the other trails, and their wakes develop under complementary flow conditions, producing the observed half-period phase lag.

The torque evolution mirrors the force behavior. During the transient, $\mathcal{T}$ rises from zero as the wakes form, then settles into quasi-periodic oscillations about a positive mean once the steady state is reached. Each oscillation corresponds to half of the orbital period, occurring at twice the orbital frequency, consistent with each BH alternately crossing its companion's wake twice per orbit. This positive mean torque reflects the wake-crossing asymmetry described in \S\ref{sec:density}: the companion's wake exerts a net gravitational pull that increases the binary's angular momentum over each orbital cycle, and the cumulative effect of these crossings produces a persistent positive torque bias. Applying the time-averaging procedure of Equation \eqref{eq:time_avg} over five orbital cycles during the steady stage ($t_1=12$ to $t_2=43.4$), we obtain
\begin{equation}
    \overline{\mathcal{T}} = 0.499 \pm 0.019,
    \label{eq:Tbar_fiducial}
\end{equation}
where the uncertainty is the standard deviation of the per-cycle averages. The ratio $\overline{\mathcal{T}} / \sigma_{\overline{\mathcal{T}}} \sim 26$ confirms that the positive torque is not a statistical fluctuation; varying the start time of the averaging window by $\pm 2$ code units changes $\overline{\mathcal{T}}$ by $\lesssim 0.02$, confirming that the mean is robust against the choice of averaging interval. Since $\overline{\mathcal{T}} > 0$, the binary gains angular momentum from the surrounding gas at $\eta=5$, implying a tendency toward orbital expansion rather than decay. This positive torque, arising from wake interaction asymmetry, is the central result of the fiducial model. Together with the density morphology presented in \S\ref{sec:density}, these results provide the baseline against which the $\eta$, $q$, and $\theta$ dependencies are measured in \S\ref{sec:parameter}.

\section{Parameter Study}\label{sec:parameter}

\subsection{Dependence on $\eta$}
\label{sec:eta_dependence}

\begin{figure*}
    \centering
    \includegraphics[width=1\linewidth]{\figdir/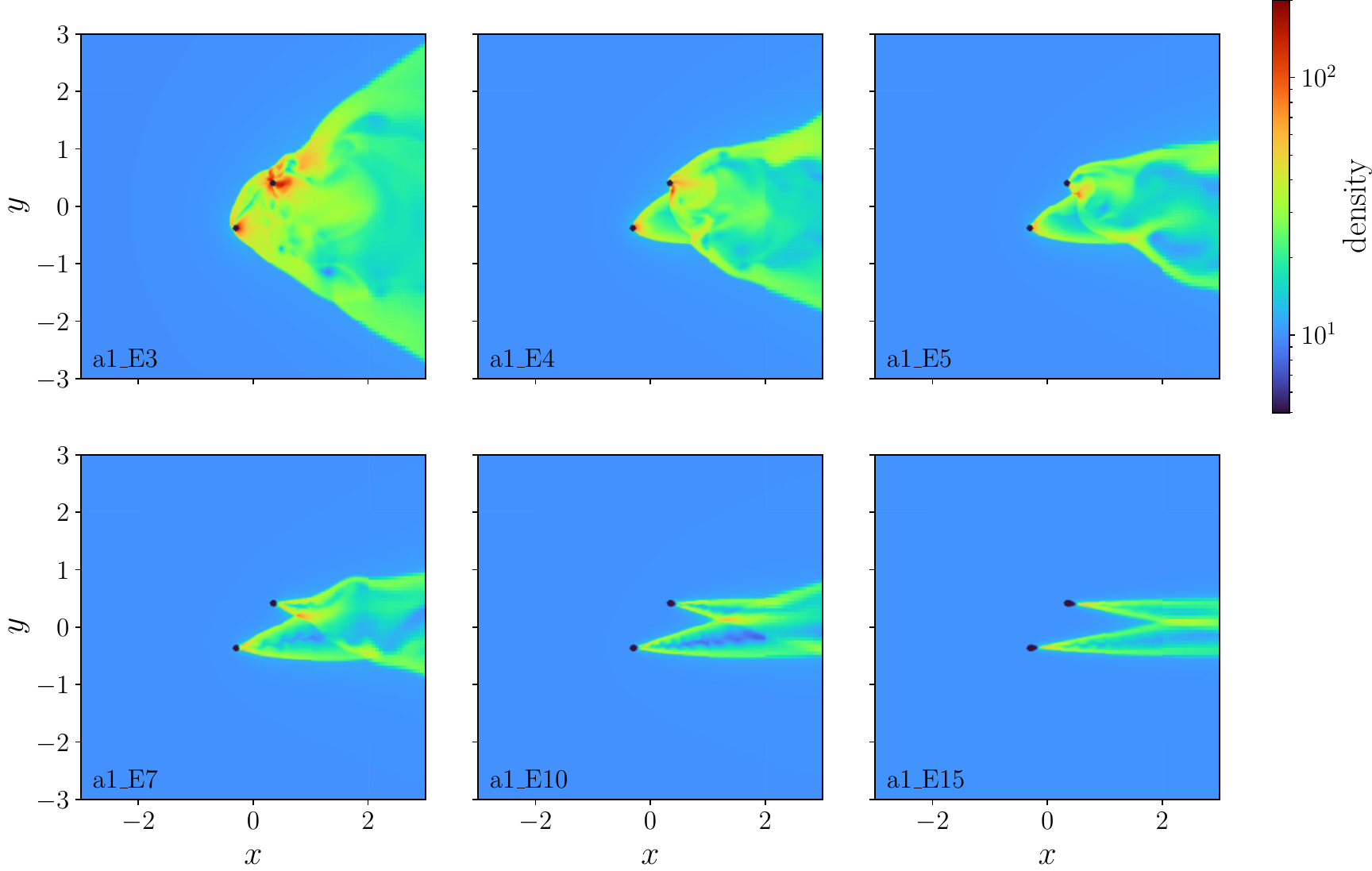}
    \caption{Gas density distributions on the $x$--$y$ plane for models with different $\eta$ values at the same simulation time ($t=48$). From left to right and top to bottom: \texttt{a1\_E3}, \texttt{a1\_E4}, \texttt{a1\_E5}, \texttt{a1\_E7}, \texttt{a1\_E10}, and \texttt{a1\_E15}.}
    \label{fig:rho_eta}
\end{figure*}

The density morphology evolves systematically with $\eta$. Figure~\ref{fig:rho_eta} shows gas density snapshots at $t=48$ for $\eta = 3, 4, 5, 7, 10, 15$. As $\eta$ increases, the wakes become progressively narrower and more spatially separated, a direct consequence of the increasing ratio $a/r_B \sim \eta^2/4$ (Equation \eqref{eq:d_over_rB}). The narrowing reflects the increasing Mach number, which constrains the wake opening angle; as $\eta$ grows, the wakes transition from broad, overlapping structures to sharply collimated Mach cones. At $\eta=7$ the wakes are clearly isolated near the binary, and at $\eta=15$ they are narrow and well separated near the center of mass, but still intersect farther downstream at the displayed time. The $\eta=5$ panel corresponds to the fiducial model presented in \S\ref{sec:fiducial}, whose separated wakes produce the positive net torque (Equation \eqref{eq:Tbar_fiducial}). At very high $\eta$, the wakes become extremely narrow and require increasingly fine spatial resolution to capture accurately; our $\eta$-series therefore extends to $\eta=20$, beyond which the wake width approaches the grid scale.

\begin{figure}
    \centering
    \includegraphics[width=1\linewidth]{\figdir/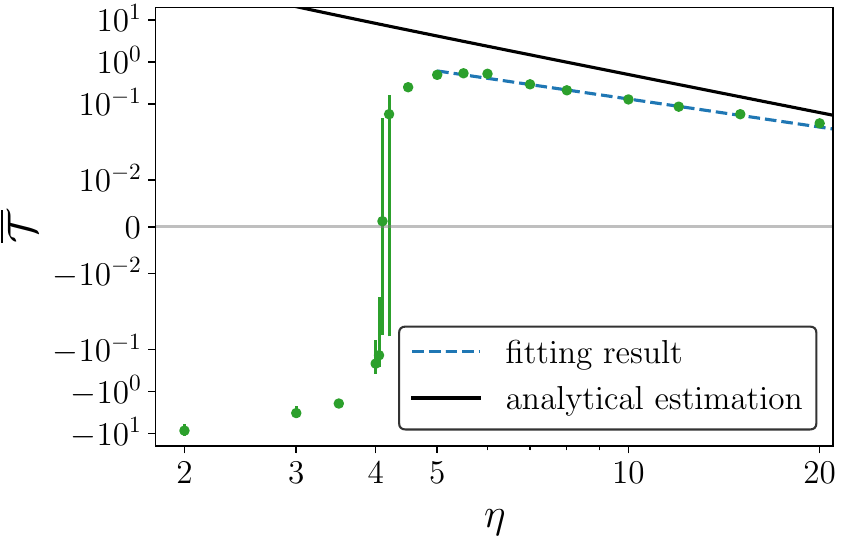}
    \caption{Time-averaged torque $\overline{\mathcal{T}}$ as a function of $\eta$ for models in the $\eta$ series. The grey line marks $\overline{\mathcal{T}}=0$. The blue dashed line shows the power-law fit $\overline{\mathcal{T}} = \mathcal{T}_0 \eta^{\alpha}$ for $\eta\geq7$. The black solid line shows the analytic prediction derived in Appendix \ref{apx:analyse}.}
    \label{fig:torq_eta}
\end{figure}

The torque responds dramatically to this morphological transition. Figure~\ref{fig:torq_eta} shows the time-averaged torque $\overline{\mathcal{T}}$ as a function of $\eta$ across the full $\eta$-series. At $\eta=2$, the torque is negative; the binary loses angular momentum to the surrounding gas (cf.\ the merged-envelope regime in \S\ref{sec:density}). The torque rises rapidly near $\eta\sim4$, crossing from negative to positive at $\eta_{\rm crit}\approx4.1$, and reaches a broad peak at $\eta\approx5$--$7$, encompassing the fiducial value. Beyond the peak, as $\eta \rightarrow \infty$, the time-averaged torque decreases monotonically toward zero. This trend is mainly because the variation of the Mach number over different orbital phases becomes increasingly negligible as the contribution of the binary orbital velocity diminishes compared to the center-of-mass velocity, making the system more symmetric. The sign change at $\eta_{\rm crit}\approx4.1$ therefore marks a critical transition: for $\eta\lesssim4.1$ the binary experiences a net torque that shrinks the orbit, while for $\eta\gtrsim4.1$ the binary gains angular momentum from the gas.

For $\eta\geq 7$, the torque follows a power-law decay,
\begin{equation}
    \overline{\mathcal{T}} = \mathcal{T}_0 \eta^{\alpha}, \qquad \alpha = -2.22 \pm 0.07,
    \label{eq:Tbar_eta_powerlaw}
\end{equation}
where the fit is performed over $\eta\in[7,20]$ and excludes the peak region ($\eta=4$--$5$) where the wake morphology transitions from overlapping to separated Mach cones. Figure~\ref{fig:torq_eta} compares this decay against the analytic prediction $\overline{\mathcal{T}}\propto(\eta^3-\eta)^{-1}$ derived in Appendix~\ref{apx:analyse}. The simulation results are systematically lower than this prediction at all $\eta$: at $\eta\gtrsim10$ the difference narrows but the analytic curve still overestimates the torque magnitude by a factor of several, while at lower $\eta$ the two diverge qualitatively: the analytic prediction remains positive and grows monotonically toward small $\eta$, whereas the simulated torque becomes negative for $\eta\lesssim4.1$ and peaks at $\eta\sim5$--$7$. These discrepancies demonstrate that the classical independent-dynamical-friction picture is unreliable for predicting the torque on a binary system. The deviation intensifies at smaller $\eta$ because the analytic model (Appendix~\ref{apx:analyse}) treats each black hole independently; it neglects the gravitational influence of each black hole on its companion's wake and the mutual attraction between the two density enhancements. In the simulations, one black hole can modify the wake of its companion as it passes through it, and the gravitational pull between each black hole and the companion's overdense wake produces an additional torque not captured by the independent-perturber picture; these effects strengthen at lower $\eta$ as broader wakes enhance both wake--BH and wake--wake interactions. %At $\eta\lesssim3$, the wakes merge into a common overdense envelope (visible in the $\eta=3$ panel of Figure~\ref{fig:rho_eta}) and the torque becomes negative, consistent with the merged-wake regime identified by \citet{2019ApJ...884...22A}.

This $\eta$-dependent transition, from merged envelope to isolated wakes, and the associated torque sign reversal, provides the framework for examining the mass ratio and inclination dependences in the following subsections.

\subsection{Dependence on Mass Ratio}
\label{sec:q_dependence}

\begin{figure}
    \centering
    \includegraphics[width=1\linewidth]{\figdir/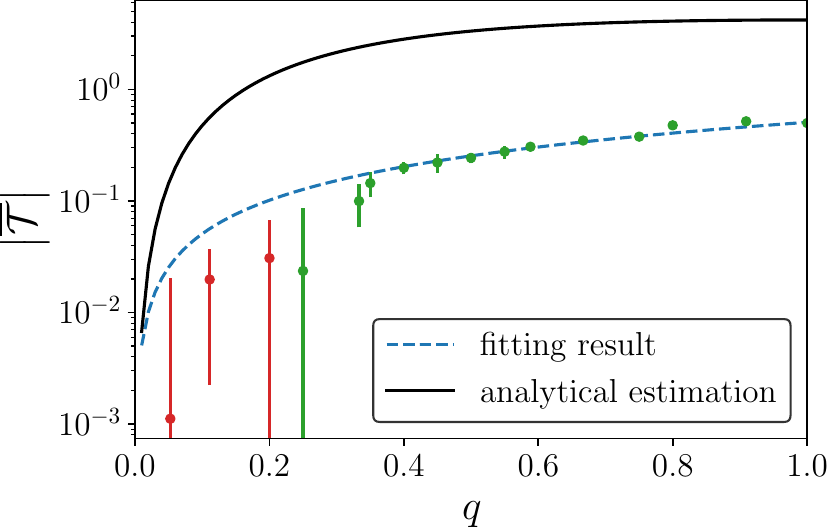}
    \caption{Absolute time-averaged torque $\left|\overline{\mathcal{T}}\right|$ as a function of mass ratio $q$ for models in the $q$ series. The blue dashed line shows the linear fit $\overline{\mathcal{T}} = \mathcal{T}_0 q$. The black solid line shows the analytic prediction derived in Appendix~\ref{apx:analyse}. Red dots indicate $\overline{\mathcal{T}}<0$ and green dots $\overline{\mathcal{T}}>0$.}
    \label{fig:torq_q}
\end{figure}

Having established the $\eta$ dependence, we now examine the dependence on the binary mass ratio $q$. Figure~\ref{fig:torq_q} shows the time-averaged torque as a function of $q$ for models at fixed $\eta=5$, with $q = 0.05$-$0.9$ and the fiducial model ($q=1$). 
%The $q=1$ case corresponds to the fiducial model of \S\ref{sec:fiducial}, for which Equation \eqref{eq:Tbar_fiducial} gives $\overline{\mathcal{T}} = 0.499 \pm 0.019$, consistent with the linear fit below within overlapping error bars. 
The torque follows a linear relation,
\begin{equation}
    \overline{\mathcal{T}} = \mathcal{T}_0 q, \qquad \mathcal{T}_0 = 0.51 \pm 0.03,
    \label{eq:Tbar_q_linear}
\end{equation}
with the data points scattering tightly around this fit and showing no systematic curvature across the tested range $q \in [0.05, 1]$. The black solid line in Figure~\ref{fig:torq_q}, showing the analytic extension to the predictions in Appendix~\ref{apx:analyse}, indicates that the simulation results are systematically lower than this prediction, an offset consistent with the discrepancy identified in \S\ref{sec:eta_dependence} — the independent-dynamical-friction picture overestimates the torque magnitude at all $\eta$ as it neglects wake--wake and wake--BH interactions. This linear scaling is consistent with the limiting behavior of the system. As $q \to 0$, the binary reduces to a single black hole interacting with the gas flow, and the net torque must vanish, which Equation \eqref{eq:Tbar_q_linear} captures naturally. At fixed total mass and orbital separation, increasing $q$ shifts more mass into the secondary component and increases the torque, implying that the wake asymmetry driving the torque grows stronger as the two components approach equality.

The linear torque scaling also allows an estimate of the orbital expansion timescale $\tau_{\rm exp} = L/\overline{\mathcal{T}}$, where $L = \mu\sqrt{GM_{\rm tot}a}$ is the orbital angular momentum of a circular binary and $\mu = M_1M_2/M_{\rm tot}$ is the reduced mass. For fixed $M_{\rm tot}$ and $a$, the reduced mass is $\mu = M_{\rm tot}\,q/(1+q)^2$, so the angular momentum scales as $L \propto q/(1+q)^2$. With $\overline{\mathcal{T}} \propto q$ from Equation \eqref{eq:Tbar_q_linear}, the timescale becomes $\tau_{\rm exp} = L/\overline{\mathcal{T}} \propto (1+q)^{-2}$. %, independent of the linear-fit constant $\mathcal{T}_0$. 
The factor $(1+q)^2$ varies from roughly unity at $q \ll 1$ to $4$ at $q=1$, so the expansion timescale varies by at most a factor of a few across the tested range — the dominant uncertainty in $\tau_{\rm exp}$ comes from the overall torque normalization rather than the $q$-scaling. The expansion timescale decreases with increasing $q$: a more equal-mass binary experiences faster orbital expansion for a given total mass and separation. That said, extrapolation to extreme mass ratios ($q \ll 0.05$) is not validated by the current data. At extreme $q$, the wake of the less massive component becomes very weak, and the torque may transition to a different scaling regime, so caution is warranted when extending these results beyond the tested range.

Together with the $\eta$-dependence established in \S\ref{sec:eta_dependence}, which fixes the overall torque magnitude at $q=1$, the $q$-scaling provides a relatively more complete description of the torque magnitude for equal-separation binaries. The $\eta$ and $q$ scalings together determine the torque, with only the geometric modulation remaining. Combining the two scalings yields $\overline{\mathcal{T}} \propto q\,\eta^{\alpha}$ for $\eta \ge 7$, where $\alpha = -2.22 \pm 0.07$ from \S\ref{sec:eta_dependence}. The orbital inclination $\theta$ introduces an additional dependence through the projection of the orbital motion onto the gas flow direction, modulating both the sign and magnitude of the torque, which we examine next.

\subsection{Dependence on Inclination Angle}
\label{sec:theta_dependence}

\begin{figure}
    \centering
    \includegraphics[width=1\linewidth]{\figdir/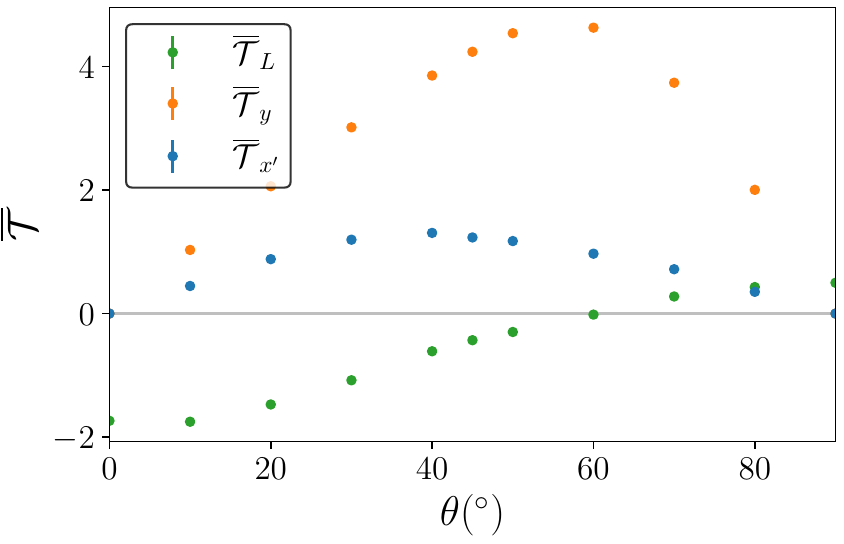}
    \caption{Time-averaged torque components $\overline{\mathcal{T}}_L$, $\overline{\mathcal{T}}_y$, and $\overline{\mathcal{T}}_{x'}$ as functions of inclination angle $\theta$ for models in the $\theta$ series at fixed $\eta=5$ and $q=1$. The grey line marks $\mathcal{T}=0$.}
    \label{fig:torque_incl}
\end{figure}

The results presented so far assume an edge-on orbit ($\theta=90^\circ$), the configuration of the fiducial model (\S\ref{sec:fiducial}), for which Equation \eqref{eq:Tbar_fiducial} gives $\overline{\mathcal{T}} = 0.499 \pm 0.019$, and accordingly reflect only one slice of the full parameter space. We now examine how the inclination angle $\theta$ between the binary's angular momentum vector and $-x$ direction modulates the torque. Figure~\ref{fig:torque_incl} shows three components of the time-averaged torque as functions of $\theta$: $\mathcal{T}_L$ (aligned with $\mathbf{L}_{\rm orb}$, governing orbital expansion or contraction), $\mathcal{T}_y$ (the $y$-component), and $\mathcal{T}_{x'}$ (directed along $\hat{x'} = \hat{y} \times \hat{L}$). We first focus on $\mathcal{T}_L$, the component that directly determines the orbital evolution. At fixed $\eta=5$, $q=1$, $e=0$, with $\theta$ spanning $0^\circ$ to $90^\circ$, $\overline{\mathcal{T}}_L$ decreases monotonically from $\overline{\mathcal{T}} \approx 0.5$ at $\theta=90^\circ$, crosses zero at $\theta \approx 60^\circ$, and reaches $\overline{\mathcal{T}} \approx -1.7$ at $\theta=0$. The torque therefore changes sign as the inclination decreases: the binary experiences orbital expansion for near-edge-on configurations ($\theta\gtrsim60^\circ$) and orbital shrinkage for near-face-on configurations ($\theta\lesssim60^\circ$). The magnitude of the face-on torque is approximately three times larger than the edge-on value, highlighting the stronger interaction when the gas flow is orthogonal to the orbital plane.

This trend follows from the geometric dependence of the gas--binary interaction. In an edge-on orbit ($\theta=90^\circ$), each black hole moves alternately downstream and upstream relative to the gas, experiencing a sinusoidal variation in its effective Mach number over one orbit. This periodic modulation produces the asymmetric wake structure identified in \S\ref{sec:density} that drives a net positive torque. As $\theta$ decreases toward $0$, the orbital plane becomes increasingly perpendicular to the gas flow, reducing the amplitude of this velocity modulation. At $\theta=0$, the gas flows orthogonal to the orbital plane: each black hole moves purely within the plane while the gas streams through in the perpendicular direction. Each black hole generates a wake that, when projected onto the orbital plane, trails opposite to its instantaneous direction of motion; the resulting dynamical friction on each component opposes the orbital motion and produces a net negative torque that is nearly independent of orbital phase. The sign reversal at $\theta \approx 60^\circ$ thus marks the balance point where the edge-on wake-crossing asymmetry (driving expansion) is compensated by the face-on projected dynamical friction (driving contraction).

The perpendicular components $\mathcal{T}_y$ and $\mathcal{T}_{x'}$ govern the orientational evolution of the binary orbit. 
$\mathcal{T}_{x'}$ drives inclination evolution via ${\rm d}\theta/{\rm d}t = \overline{\mathcal{T}}_{x'}/L_{\rm orb}$. For $0<\theta<90^\circ$, $\overline{\mathcal{T}}_{x'} > 0$ drives $\theta$ toward $90^\circ$, indicating that near-edge-on configurations are attractors for the inclination evolution; at $\theta = 90^\circ$, $\overline{\mathcal{T}}_{x'} \sim 0$ and inclination precession ceases. The magnitudes of all three components are comparable ($\sim O(1)$ in code units), implying that the timescales for orbital expansion or contraction, precession, and inclination evolution are all similar.%, and all are substantially shorter than the gravitational-wave inspiral timescale for the orbital separations considered here.

Together with the $\eta$- and $q$-dependences established in \S\ref{sec:eta_dependence} and \S\ref{sec:q_dependence}, this completes the parametric characterization of the gas torque on the binary. The three parameters control distinct aspects of the interaction: $\eta$ governs the sign and overall magnitude through the wake morphology and the degree of wake overlap (\S\ref{sec:eta_dependence}), $q$ sets the linear scaling with mass ratio, and $\theta$ modulates both the magnitude and sign through the orbital geometry, with the torque crossing from positive to negative at $\theta \approx 60^\circ$. 
%These three dependences together demonstrate that the gas torque on a circular binary is a well-defined function $\overline{\mathcal{T}}(\eta, q, \theta)$ with a predictable sign and magnitude over the explored parameter space. 
These three dependences, each measured along a one-dimensional slice of the parameter space, suggest that the gas torque on a circular binary is approximately separable as $\overline{\mathcal{T}}\approx \overline{\mathcal{T}}_0f_{\eta}(\eta)f_q(q)f_{\theta}(\theta)$ over the surveyed range, where $f_{\eta}(\eta)$, $f_q(q)$ and $f_{\theta}(\theta)$ describe the individual scalings reported above. 
In particular, the identification of a large-$\eta$, edge-on regime in which $\overline{\mathcal{T}} > 0$, bounded by $\eta \gtrsim \eta_{\rm crit} \approx 4.1$ and $\theta \gtrsim \theta_{\rm crit} \approx 60^\circ$, suggests that gas-embedded BBHs may undergo orbital expansion under astrophysically relevant conditions. Having established these dependences, we now discuss the implications for binary black holes in AGN disk environments.

\section{Discussion}\label{sec:discussion}

\subsection{Comparison to Prior Work}
\label{sec:comparison}

The binary BHL accretion flow computed by \citet{2019ApJ...884...22A} at ${\cal M}_\infty=2$ and $a_0/R_a=1$ corresponds to $\eta=2$ in our parameterization. Our \texttt{a1\_E2} simulation reproduces its density distribution, confirming that the merged-envelope regime ($\eta < \eta_{\rm crit}$) encompasses the conditions they explored. The agreement extends to the torque magnitude: applying the scaling relation of \citet{2019ApJ...884...22A} (their Equation~(52)) to the AGN disk parameters of \S\ref{sec:disc_agn} ($R=0.01$~pc, $M_1=M_2=10\,M_\odot$, $a=2$~AU) gives an inspiral timescale $\tau_{\rm insp,gas} \equiv a/\dot{a} \sim 7\times10^2$~yr for $\eta=2$, consistent with $\tau_{\rm insp,gas}' \sim 6\times10^2$~yr obtained from the time-averaged torque of our \texttt{a1\_E2} model. \citet{2019ApJ...884...22A} focused on BHL mass accretion rates and morphological characterization of the shared-envelope structure but did not isolate $\eta$ as the governing dimensionless parameter. Our work extends this regime to $\eta = 3$--$20$ and reveals the torque sign reversal at $\eta_{\rm crit}\approx4.1$. The consistency of our $\eta=2$ results with their published density snapshots and torque estimates demonstrates code agreement in the overlapping regime, lending confidence to our higher-$\eta$ results.

\citet{2022ApJ...940..155D} identified a separation-dependent torque sign reversal in 3D shearing-box simulations of BBHs in AGN disks: binaries at sufficiently small separations undergo expansion, while those at larger separations contract. Their mechanism is governed by the ratio of binary separation to disk scale height ($a/H$), with the transition occurring when the separation exceeds the scale height and the circumbinary disk breaks into individual circum-single disks. This is physically distinct from our $\eta$-based mechanism: in \citet{2022ApJ...940..155D}, the torque sign depends on the structure of circum-single and circumbinary disks formed in a Keplerian shear flow, whereas in our work it depends on $v_g/v_o$ in a uniform flow. The CSD-based simulations of \citet{2021ApJ...911..124L,2022MNRAS.517.1602L} and the shearing-box study of \citet{2022ApJ...940..155D} operate at small $v_g/v_o$, where individual circum-single disks can form; our results apply at large $v_g/v_o$, where CSDs cannot form and the wake interaction dominates the torque. These two mechanisms may operate simultaneously in realistic AGN disks; even in regions where $a/H \gg 1$ would warrant contraction in the picture of \citet{2022ApJ...940..155D}, a sufficiently large $\eta$ could still produce a net expansion torque.

The independent dynamical friction picture treats each BH as a separate perturber and sums the torque contributions linearly. Appendix~\ref{apx:analyse} demonstrates that this approach fails quantitatively for the binary: it produces torque magnitudes systematically exceeding the simulation results by factors of several
found in our simulations (\S\ref{sec:eta_dependence}). The discrepancy arises from the wake--wake and wake--BH gravitational coupling, absent from single-perturber theory. Together, these results situate the binary wake torque $\overline{\mathcal{T}}(\eta)$ as bridging the gap between the merged-envelope regime of \citet{2019ApJ...884...22A} and the independent-DF limit for well-separated single perturbers.

\subsection{Implications for BBH Mergers in AGN Disks}
\label{sec:disc_agn}

We adopt a $\beta$-disk model \citep{2011PhRvD..84b4032K} with the parameters of \citet{2017ApJ...835..165B} ($\alpha=0.3$, $\dot{M}_\bullet = 0.1\,\dot{M}_{\bullet,\rm Edd}$, $M_\bullet = 10^6\,M_\odot$) at $R = 0.01$~pc. The disk scale height, density, and radial inflow velocity are
\begin{equation}
\begin{split}
    H&\approx 10^{14}~{\rm cm}~\left(\frac{R}{0.01~\rm pc}\right)^{-3/5} \left(\frac{M_{\bullet}}{10^6~M_{\odot}}\right)^{4/5},\\
    \rho&\approx 6\times 10^{12}m_p~{\rm cm^{-3}}~\left(\frac{R}{0.01~\rm pc}\right)^{-33/20} \left(\frac{M_{\bullet}}{10^6~M_{\odot}}\right)^{19/20},\\
    v_r^{\rm gas}&\approx 0.006~{\rm km\,s^{-1}}\left(\frac{R}{0.01~{\rm pc}}\right)^{-2/5}\left(\frac{M_{\bullet}}{10^6~M_\odot}\right)^{1/5}.
\end{split}
\end{equation}
For a BBH with each component of $10\,M_\odot$ and separation $a=2$~AU, $v_o \sim 47~\rm km\,s^{-1}$. At these parameters, the simulation results of fiducial model yield an orbital expansion timescale $\tau_{\rm exp} = L/\overline{\mathcal{T}} \sim 7\times10^3$~yr at $R=0.01$~pc. Since $\rho \propto R^{-33/20}$ and $\tau_{\rm exp} \propto \rho^{-1}$, the expansion timescale scales as $\tau_{\rm exp} \propto R^{33/20}$, reaching $\sim10^6$~yr at $R \sim 0.2$~pc, shorter than AGN disk lifetimes ($\sim10^7$--$10^8$~yr; \citealt{2017ApJ...835..165B}).

Three main channels produce BBHs embedded in AGN disks: (i) \emph{in-situ} formation via disk fragmentation at parsec scales \citep{2017MNRAS.464..946S}, (ii) capture of nuclear cluster binaries crossing the disk on inclined orbits \citep{2017ApJ...835..165B}, and (iii) binary accumulation at migration traps at $20$--$300\,R_{\rm sch}$ \citep{2016ApJ...819L..17B}. For channels (i) and (iii), the BBH co-rotates with the disk gas, giving $v_g \simeq v_r^{\rm gas} \ll v_o$ and $\eta \ll 1$, so the time-averaged torque is negative. For channel (ii), the CM velocity relative to the disk gas is set by the outer orbital inclination $\psi$, yielding $\Delta v_{\rm CM} \simeq v_K \sin\psi$ and $\eta \gtrsim 1$ for $\psi \gtrsim$ a few degrees, naturally accessing the positive-torque regime. Since $c_s \simeq 2~\rm km\,s^{-1}$ at $R=0.01$~pc, $\Delta v_{\rm CM}$ also exceeds the sound speed of the disk gas at small inclinations $\psi$. This is in contrast to most prior studies of gas-embedded binaries, which operate in the subsonic regime ($\Delta v_{\rm CM}<c_s$), and motivates our supersonic simulations, whose flow Mach number $\mathcal{M}=v_g/c_s$ spans $2$--$20$.

\citet{2017ApJ...835..165B} find that only $\sim20\%$ of captured binaries align with the disk within $10^8$~yr. While $\psi > \psi_{\rm crit}$, where
\begin{equation}
v_K\sqrt{(1-\cos\psi_{\rm crit})^2+\sin^2\psi_{\rm crit}} = v_o\eta_{\rm crit},
\end{equation}
giving $\psi_{\rm crit} \simeq 17^\circ$ for $\eta_{\rm crit}=4.1$ at $R=0.01$~pc, the binary experiences positive torque and expands. For $\psi = \pi/2$, the cumulative residence time within the disk before alignment is at least $(H/R)\,\tau_{\rm align} \sim 10^5$~yr, exceeding $\tau_{\rm exp} \sim 7\times10^3$~yr by more than an order of magnitude.

Assuming an isotropic distribution of inner orbit inclinations $\theta$, the positive-torque window ($\theta \gtrsim 60^\circ$, from \S\ref{sec:theta_dependence} and Figure~\ref{fig:torque_incl}) extends to $60^\circ \lesssim \theta \lesssim 120^\circ$ under the symmetry $\theta \leftrightarrow 180^\circ-\theta$, which follows from the invariance of the physical setup under reversal of the orbital angular momentum direction while holding the gas flow fixed. For an isotropic distribution this window contains $\int_{60^\circ}^{120^\circ}\sin\theta\,d\theta / \int_{0^\circ}^{180^\circ}\sin\theta\,d\theta = 1/2$ of the inner orbits. Moreover, the inclination precession discussed in \S\ref{sec:theta_dependence}, where $\overline{\mathcal{T}}_{x'} > 0$ drives $\theta$ toward $90^\circ$, pushes binaries initially in the negative-torque window toward the edge-on configuration, so that a larger fraction of the population eventually enters the positive-torque regime than the instantaneous $\sim50\%$ estimate. For these systems, $\tau_{\rm exp} \ll \tau_{\rm align}$ implies that binary separations increase by factors of several before alignment completes, significantly lengthening the post-alignment GW inspiral timescale and suppressing the merger rate from this channel. Our results therefore suggest that the predicted rate may require downward revision, though quantification demands a population synthesis calculation.

LISA may detect gas-induced orbital evolution in BBHs prior to LIGO/Virgo detection \citep{2021PhRvL.126j1105T}; the sign of the frequency drift could provide a direct observational test of the expansion versus inspiral regimes mapped by $\eta$.

\subsection{Caveats and Future Work}
\label{sec:caveats}

Several approximations in our numerical setup affect the quantitative applicability of these results. (i) Uniform medium: realistic AGN disks possess vertical density stratification and radial gradients, which may modify the wake opening angle and shock structure. Where the gradient scale is much larger than $d$, the uniform-medium approximation is valid; near sharp density jumps, the torque may deviate from our predictions. (ii) Fixed orbit: our simulations hold the binary orbit fixed, isolating the instantaneous gas torque from orbital back-reaction. In reality, a positive torque increases the separation, which decreases $v_o$ ($\propto a^{-1/2}$) and increases $\eta$; this constitutes a feedback loop that could amplify or stall the expansion. (iii) No MHD: magnetic fields may provide additional angular momentum transport and energy dissipation channels not captured in our purely hydrodynamic treatment. (iv) Circular orbits: eccentric binaries experience time-dependent $v_g/v_o$ and episodic disk crossings, with possible secular effects explored by \citet{2024ApJ...970..107C}. (v) Adiabatic index: all simulations adopt $\gamma=5/3$. The wake structure, Mach cone opening angle, and post-shock density enhancement depend on the gas equation of state; a softer equation of state (smaller $\gamma$) yields stronger shock compression, while a stiffer one tends to weaken it. The quantitative torque values and the precise location of $\eta_{\rm crit}$ may therefore shift for other $\gamma$, although the qualitative sign reversal between merged and separated wakes should persist. Systematic exploration of the $\gamma$ dependence is left to future work.

These limitations suggest several directions for future investigation. (a) Live orbital evolution incorporating gas torque feedback would determine whether the expansion reaches a terminal radius or continues indefinitely. (b) Simulations in stratified, shearing AGN disk backgrounds would reconcile our $\eta$-dependent results with the $a/H$-dependent mechanism of \citet{2022ApJ...940..155D}, potentially revealing a general two-parameter ($\eta$, $a/H$) torque map. (c) Extension to eccentric and inclined outer orbits would cover the full dynamical phase space relevant to AGN disk capture scenarios. (d) Population synthesis incorporating the $\eta$-dependent torque prescription from~\S\,5 could quantify the merger rate suppression estimated in~\S\,6.3. (e) Our simulations assume isolated BHL accretion without outflows. In many astrophysical binaries, however, one or both components drive winds or jets. \citet{2020MNRAS.494.2327L} and \citet{2025ApJ...992....7L} demonstrated that outflows can produce accelerating forces (anti-friction) on single objects, and \citet{2022ApJ...932..108W} showed that colliding outflows in a binary generate a positive torque through a mechanism distinct from the wake interaction studied here. How outflows modify the wake-driven torque, and whether they enhance or disrupt the expansion regime we identify, is beyond the scope of the present work and merits dedicated investigation.

%We note that \citet{2025ApJ...992....7L} identified an equilibrium radius near $4000\,R_{\rm sch}$ for a $10^8\,M_\odot$ SMBH where single objects accumulate via dynamical friction--antifriction balance; binaries formed there would naturally satisfy $\eta \gtrsim 4.1$, providing an independent channel for the expansion regime identified here.

Despite these approximations, the fundamental result, that $\eta$ governs the gas torque sign and magnitude via wake interactions, is robust: the physics of separated vs.\ merged wakes does not depend on the disk microphysics, only on the velocity ratio. The scaling relations established here therefore provide a foundation on which higher-fidelity treatments can be built.

\section{Conclusions} \label{sec:conclusion}

We perform hydrodynamic simulations of binary black holes moving through a uniform gaseous medium and study the gravitational drag and torque generated by gas wake interaction. The main results are summarized as follows. 
\begin{enumerate} 
\item The dimensionless velocity ratio $\eta = v_g/v_o$ governs both the gas wake morphology and the net torque on the binary. At $\eta \lesssim 3$, the two wakes merge into a shared overdense envelope; at $\eta \gtrsim 5$, they form narrow, separated Mach cones that periodically cross each other's trajectories. 
\item A critical value $\eta_{\rm crit} \approx 4.1$ separates the negative-torque ($\eta < \eta_{\rm crit}$) from the positive-torque ($\eta > \eta_{\rm crit}$) regime. For the fiducial model ($\eta=5$, $q=1$, $\theta=90^\circ$), the time-averaged torque $\overline{\mathcal{T}} = 0.499 \pm 0.019$ is positive at $>25\sigma$, confirming that the binary gains angular momentum from the gas. 
\item The torque arises from gravitational coupling between each black hole and the density wake produced by its companion, a wake-interaction mechanism fundamentally distinct from independent dynamical friction. Appendix~\ref{apx:analyse} confirms that a single orbiting particle reproduces the classical DF prediction, whereas the binary torque deviates because the wakes are coherent structures that couple the two BHs gravitationally. 
\item The torque scales systematically with the governing parameters. For $\eta \geq 7$, $\overline{\mathcal{T}} \propto \eta^{-2.22 \pm 0.07}$; the mass-ratio dependence is linear, $\overline{\mathcal{T}} = \mathcal{T}_0 q$ with $\mathcal{T}_0 = 0.51 \pm 0.03$; and the inclination-dependent torque changes sign near $\theta \approx 60^\circ$, with ${\sim}50\%$ of isotropic inner orbits experiencing positive torque. 
\item Under a disk model \citep{2011PhRvD..84b4032K}, $\eta \gtrsim 5$ is naturally achieved for capture-channel binaries in AGN disks. The orbital expansion timescale ranges from $\tau_{\rm exp} \sim 7 \times 10^3$~yr at $R = 0.01$~pc to ${\sim}10^6$~yr at $R \sim 0.2$~pc ($\tau_{\rm exp} \propto R^{33/20}$), shorter than AGN lifetimes (${\sim}10^7$--$10^8$~yr), suggesting that gas-driven expansion can compete with gravitational-wave inspiral and suppress the BBH merger rate from the capture channel. 
\end{enumerate}

\begin{acknowledgements}
\noindent
We acknowledge the computational resources provided by the Kavli Institute for Astronomy and Astrophysics in Peking University. We thank Xian Chen for helpful discussions. R.L. acknowledges support from the HeisingSimons Foundation 51 Pegasi b Fellowship.
\end{acknowledgements}

\appendix

\section{Comparison between Numerical Simulations and Analytic Predictions}\label{apx:analyse}

\subsection{Analytic Estimates of Dynamical Friction}

\begin{figure}
    \centering
    \includegraphics[width=0.6\linewidth]{\figdir/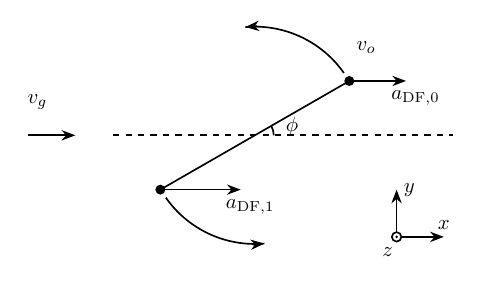}
    \caption{Schematic illustration of the analytic model in the $x$--$y$ plane. The inclination angle is $\theta=90^\circ$ (edge-on configuration).}
    \label{fig:model}
\end{figure}

Consider Ostriker's dynamical friction force \citep{1999ApJ...513..252O} in edge-on configuration ($\theta=90^\circ$), as shown in Figure \ref{fig:model}:
$$a_{DF,0}=\frac{4\pi G^2\rho M}{v_g^2+v_o^2+2v_gv_o\sin{\phi}}\mathcal{I},$$
$$a_{DF,1}=\frac{4\pi G^2\rho M}{v_g^2+v_o^2-2v_gv_o\sin{\phi}}\mathcal{I},$$
where $M$ is the mass of each particle and the mass ratio is $q=1$. $\mathcal{I}=\ln{\left(\Lambda (1-1/\mathcal{E}^2)^{1/2}\right)}$, where $\mathcal{E}=\frac{\sqrt{v_g^2+v_o^2\pm 2v_gv_o\sin{\phi}}}{c_s}$. We assume $\mathcal{I}$ is constant. 
The total torque at orbital phase $\phi$ is
$$\mathcal{T}(\phi)=-a_{DF,0}M\frac{d}{2}\sin{\phi}+a_{DF,1}M\frac{d}{2}\sin{\phi}$$
$$=\frac{2\pi G^2M^2\rho d\mathcal{I}}{\eta v_o^2} \frac{\sin^2{\phi}}{\frac{(1+\eta^2)^2}{4\eta^2}-\sin^2{\phi}}.$$

As $\eta>1$, $\frac{(1+\eta^2)^2}{4\eta^2}>1$,
$$\overline{\mathcal{T}}=\frac{1}{\pi} \int_0^{\pi}\mathcal{T}(\phi)d\phi$$
$$=\frac{4\pi G^2 M^2 \rho d\mathcal{I}}{\eta v_o^2} \frac{1}{\eta^2-1}.$$
For fixed $v_o$, $\overline{\mathcal{T}}$ can be expressed as a function of $\eta$:
$$\overline{\mathcal{T}}=\frac{4\pi G^2M^2\rho d\mathcal{I}}{v_o^2}\mathcal{E},~\mathcal{E}=\frac{1}{\eta^3-\eta}.$$

\subsection{Comparison of Analytic Estimates with Simulation Results}
Assume a single particle undergoing circular motion with orbital radius $r=d/2$. The particle mass is $M$ and the orbital velocity is $v_o$. The time-averaged torque is
\begin{equation}\label{eq:sin_torq}
    \overline{\mathcal{T}}=\frac{2\pi G^2M^2\rho d\mathcal{I}}{v_o^2}\frac{1}{\eta^3-\eta},~\eta = v_g/v_o.
\end{equation}

\begin{figure}
    \centering
    \includegraphics[width=1\linewidth]{\figdir/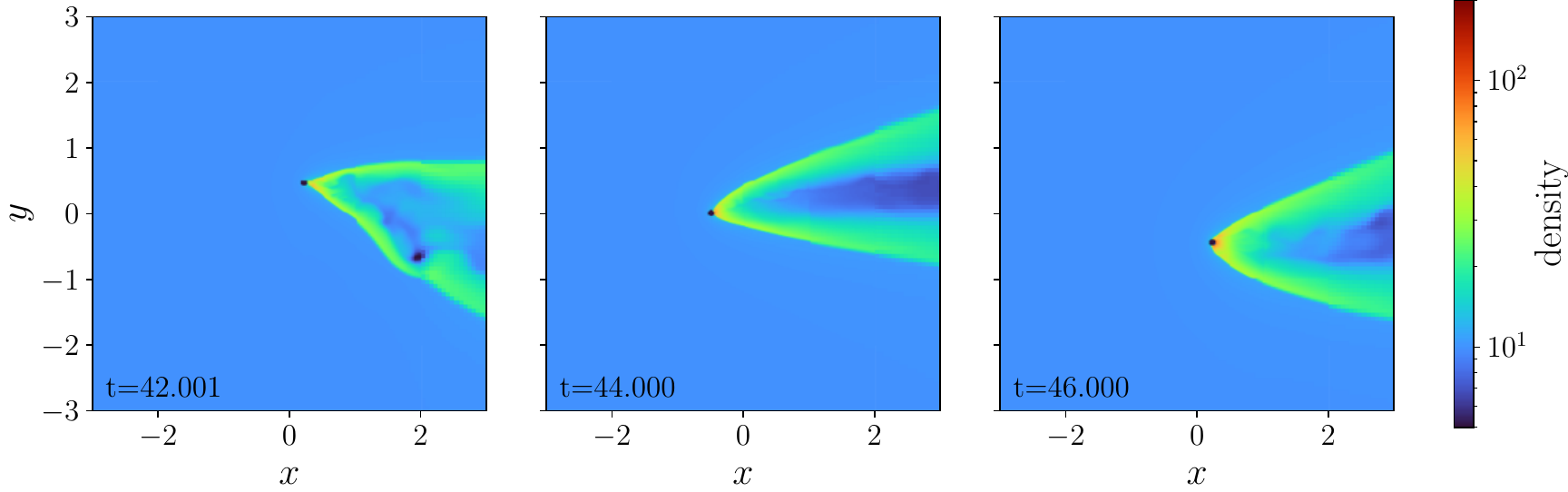}
    \caption{Gas density distribution for the single-particle model \texttt{a1\_E5\_sin}. The setup is similar to model \texttt{a1\_E5} but contains only one BH on a circular orbit.}
    \label{fig:M5_sin}
\end{figure}

We perform a simulation of a single particle in circular motion, similar to \texttt{a1\_E5} but with only one black hole (model \texttt{a1\_E5\_sin}), as shown in Figure \ref{fig:M5_sin}.

\begin{figure}
    \centering
    \includegraphics[width=1\linewidth]{\figdir/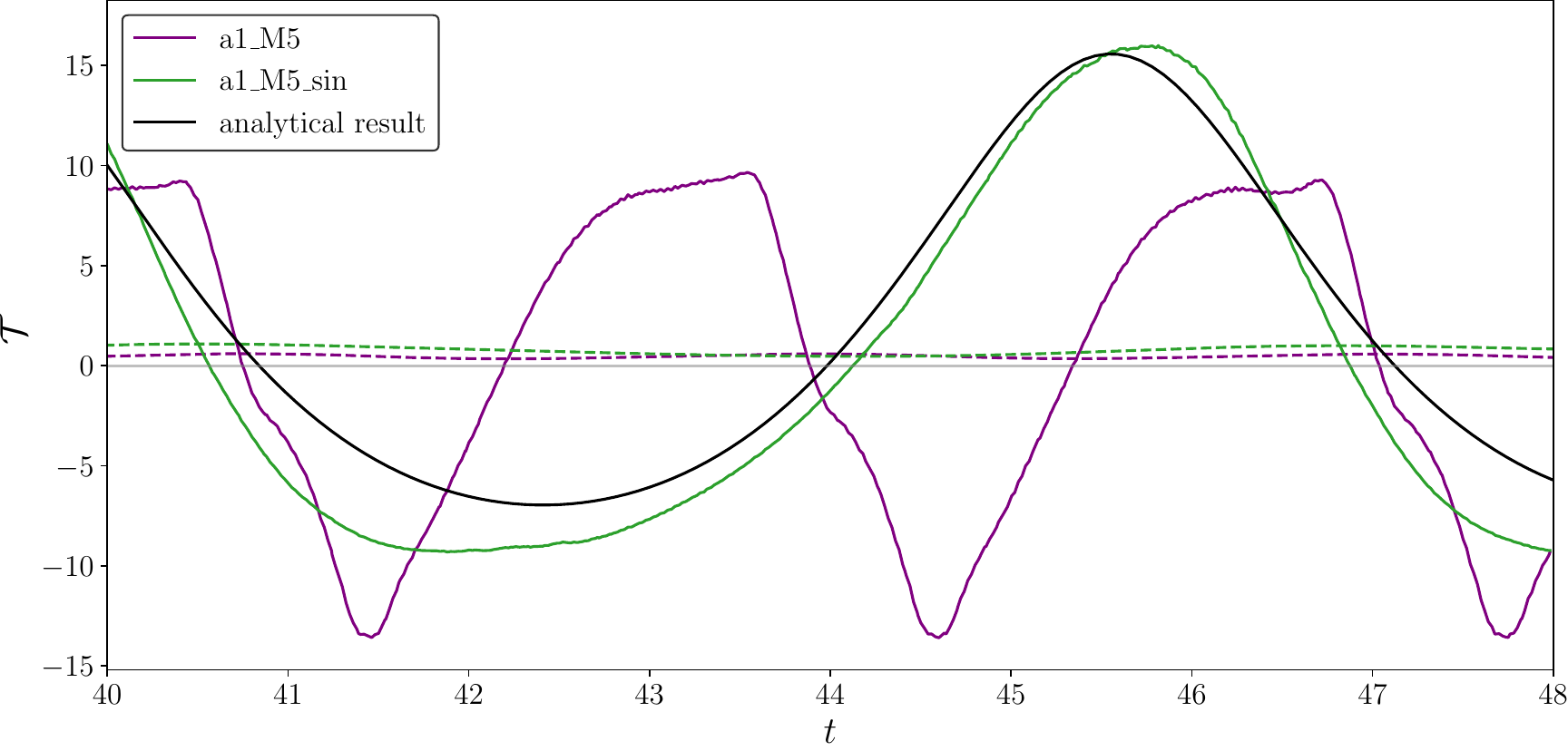}
    \caption{Comparison of torque evolution for the binary model \texttt{a1\_E5}, the single-particle model \texttt{a1\_E5\_sin}, and the analytic prediction from Equation \eqref{eq:sin_torq}. The dashed line shows the mean torque averaged from $t=0$, while the grey line marks $\mathcal{T}=0$.}
    \label{fig:torq_sin_compare}
\end{figure}

The torque curves of \texttt{a1\_E5}, \texttt{a1\_E5\_sin}, and the analytic prediction from Equation \eqref{eq:sin_torq} are shown in Figure \ref{fig:torq_sin_compare}. The \texttt{a1\_E5\_sin} model is consistent with Equation \eqref{eq:sin_torq}, while the binary case deviates from the analytic prediction. One possible reason is that the wake generated by each BH exerts an additional gravitational force on the companion BH, so the torque estimated from independent DF forces is no longer applicable.

\section{Resolution and Convergence Tests}\label{apx:convergence}

\begin{figure}
    \centering
    \includegraphics[width=1\linewidth]{\figdir/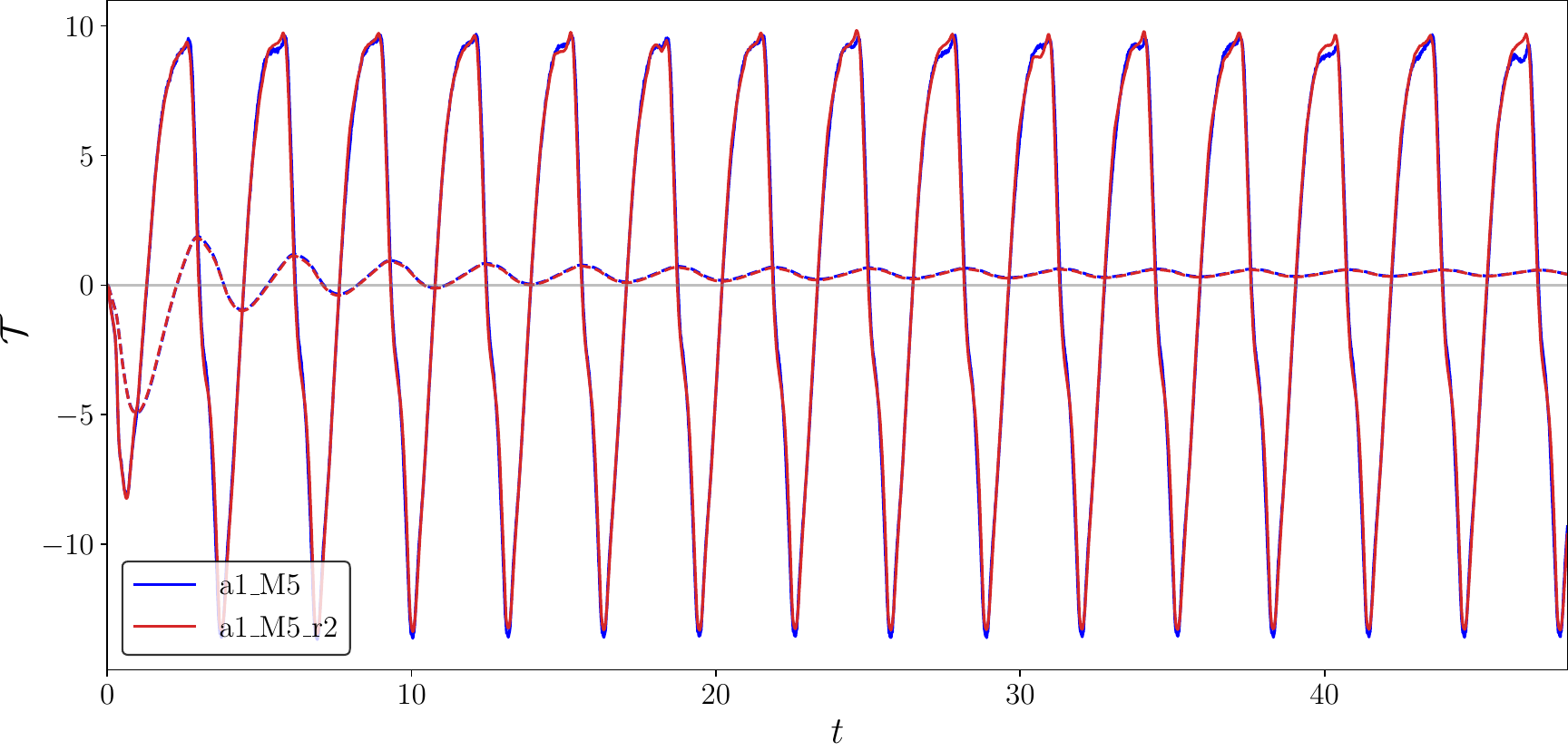}
    \caption{Comparison of torque evolution between the fiducial model \texttt{a1\_E5} and the higher-resolution model \texttt{a1\_E5\_r2}. The spatial resolution of \texttt{a1\_E5\_r2} is twice that of \texttt{a1\_E5}. The dashed line shows the mean torque averaged from $t=0$, while the grey line marks $\mathcal{T}=0$.}
    \label{fig:resolution_curve}
\end{figure}

%Compare the \texttt{a1\_E5} model and the \texttt{a1\_E5\_r2} model. The latter has twice the spatial resolution of the former. The torque curves are shown in Figure \ref{fig:resolution_curve}. In the \texttt{a1\_E5} model $\overline{\mathcal{T}}=0.499$ while in the \texttt{a1\_E5\_r2} model $\overline{\mathcal{T}}=0.528$.
All simulations in this work are carried out at the fiducial resolution described in \S\ref{sec:config}. To test numerical convergence, we compare the fiducial model \texttt{a1\_E5} with the higher-resolution run \texttt{a1\_E5\_r2}, which has twice the spatial resolution; the torque time series of the two runs are shown in Figure~\ref{fig:resolution_curve}. The time-averaged torques are $\overline{\mathcal{T}}=0.499$ for \texttt{a1\_E5} and $\overline{\mathcal{T}}=0.528$ for \texttt{a1\_E5\_r2}. The small difference shows that the fiducial resolution is sufficient and that the sign and qualitative conclusions of this work are robust to spatial resolution.

\bibliography{sample701}{}
\bibliographystyle{aasjournalv7}

\end{document}